\documentclass[namedreferences]{solarphysics}
\usepackage[hyperref,optionalrh,solaromanenum]{spr-sola-addons} % For Solar Physics 
\usepackage{graphicx}                    % For eps figures, newer & more powerfull
\usepackage{color}                       % For color text: \color command
\usepackage{breakurl}                         % For breaking URLs easily trough lines
\newcommand{\ion}[2]{#1\,{\sc #2}}
\begin{document}

\begin{article}

\begin{opening}

\title{Wave-like Formation of Hot Loop Arcades}

\author{A.~\surname{Reva}$^{1}$\sep
        S.~\surname{Shestov}$^{1}$\sep
        I.~\surname{Zimovets}$^{2}$\sep 
        S.~\surname{Bogachev}$^{1}$\sep
        S.~\surname{Kuzin}$^{1}$     
       }
       
\institute{$^{1}$ Lebedev Physical Institute, Russian Academy of Sciences
                     email: \url{reva.antoine@gmail.com}\\ 
           $^{2}$ Space Research Institute, Profsoyuznaya 84/32, Moscow 117997, Russia
             }
             
\begin{abstract}
We present observations of hot arcades made with the \ion{Mg}{xii} spectroheliograph onboard the CORONAS-F mission,
which provides monochromatic images of hot plasma in the \ion{Mg}{xii}
8.42~\AA\ resonance line. The arcades were observed to form above the polarity inversion
line between Active Regions NOAA 09847 and 09848 at four successive episodes: at 09:18, 14:13, and 22:28 UT on
28 February 2002, and at 00:40 UT on 1 March 2002. The arcades all evolved in the same way: a) a small flare (precursor) appeared near the edge of the
still invisible arcade, b) the arcade brightened in a wave-like manner~---~closer
loops brightened earlier, and c) the arcade intensity gradually decreased in
$\approx$~1~h. The estimated wave speed was $\approx$~700~km~s$^{-1}$, and the distance between the hot loops was $\approx$~50~Mm. The arcades formed without visible changes in their magnetic structure. The arcades were probably heated up by the instabilities of the current sheet above the arcade, 
which were caused by an MHD wave excited by the precursor.
\end{abstract}
\keywords{Flares, Waves}
\end{opening}

\section{Introduction}

It is generally accepted that the energy source of solar flares is the energy of  the coronal
magnetic field.  According to the standard flare model \citep{Carmichael1964, Sturrock1966, Hirayama1974, Kopp1976}, a magnetic reconnection occurs in
the corona. Inside the reconnection region, electrons are accelerated.  The accelerated electrons propagate to the lower layers of the solar atmosphere,  stop in the chromosphere, heat it, and create hard X-ray (HXR) sources. The heated chromoshperic material expands into the corona filling the magnetic field lines with hot and dense plasma. Many flares show signs of the standard model: HXR emission at the loop footpoints and apexes, plasma upflows inferred from Doppler shifts, and magnetic field configuration highlighted in extreme ultra-violet (EUV) images.

Post-flare arcades are a good testing ground for the standard flare model. Whereas arcades are frequently observed with EUV telescopes, a `truly hot' arcade is not as frequent. In fact, the only report we found is given by \citet{Warren1999}. Slightly more numerous are observations of HXR ribbons in the arcade footpoints \citep{Masuda2001, Liu2007, Jing2007, Krucker2011}. In all these articles, the authors stressed that the properties of the phenomena under analysis ---  H$_\alpha$ and HXR ribbons, magnetic reconfiguration, sheared magnetic field, and heating and cooling of plasmas --- are consistent with the standard flare model.

The standard flare model is 2D or 2.5D in nature. However, some observations require a 3D explanation; for example, sequential brightening of the flare arcade \citep{Vorpahl76}, spreading of heating from a localized flare to a neighboring active region \citep{Parenti2010}, HXR footpoint motion along the arcade \citep{Bogachev2005, Grigis2005}, and asymmetric filament eruption \citep{Liu2009, Tripathi2006}. To further develop a flare theory, we need to extend the standard flare model into 3D.

It is widely accepted that changes in the coronal magnetic structure~---~flux emergence, flux cancellation, or footpoints shearing motion --- may drive flare reconnection \citep[see reviews such as][]{Benz2008, Fletcher2011}. However, several authors have proposed that MHD waves can trigger instabilities in the current sheet, which will lead to the reconnection \citep{Vorpahl76, Somov1982, Nakariakov2006, McLaughlin2011, Nakariakov2011, Artemyev2012}. 

To observationally investigate flare heating, we need some way to see  hot plasma. The most commonly used instruments for hot plasma research --- the {\it Atmospheric Imaging Assembly} \citep[AIA;][]{Lemen2011} onboard the {\it Solar Dynamic Observatory (SDO)}, the {\it X-Ray Telescope} \citep[XRT;][]{Golub2007} onboard {\it Hinode}, and the {\it Soft X-ray Telescope} \citep[SXT;][]{Tsuneta1991} onboard {\it Yohkoh} --- have wide spectral sensitivity. Their images contain cold plasma background, and isolation of hot plasma requires various reasonable assumptions \citep{Reale2011, Warren2012, Testa2012}. It would be ideal for flare heating research to obtain monochromatic images of the solar corona in a hot monochromatic line.

In this work, we present the first observations of the formation of hot loop arcades in a hot monochromatic line. The behavior of the observed arcades cannot be explained with the 2.5D standard flare model: they formed in a wave-like manner without changes in their magnetic structure. The aim of this work is to present the observations and give them a reasonable interpretation.

\section{Observational Data}
We analyzed hot loop arcades observed on 28 February 2002 at 09:18, 14:13, and 22:28 UT and on 1 March 2002 at 00:40 UT. We mainly used the \ion{Mg}{xii} spectroheliograph \citep{zhi03a} data to investigate the behavior of hot arcades, the {\it Michelson Doppler Imager} \citep[MDI;][]{Scherrer1995} data  to infer the photospheric magnetic field beneath the arcades, the {\it Reuven Ramaty High Energy Solar Spectroscopic Imager} \citep[RHESSI;][]{Lin2002} data to determine where the accelerated particles were present in the arcade, and the {\it Extreme ultraviolet Imaging Telescope} \citep[EIT;][]{del95} data  to see how the arcades looked like in low-temperature lines.

\begin{figure}[t]
\centering
\includegraphics[width = \textwidth]{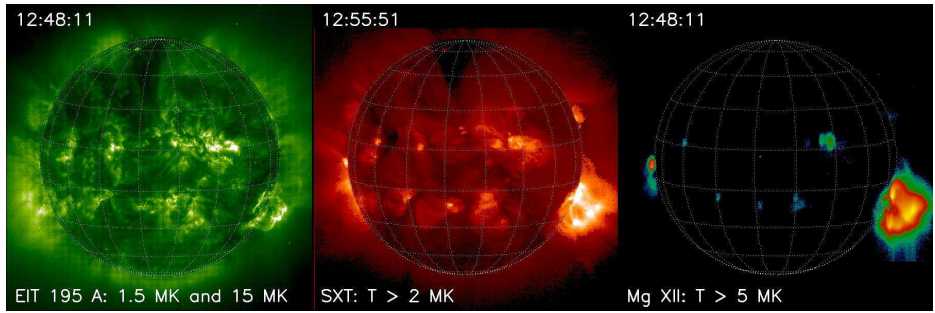}
\caption{Comparison of hot plasma imagers. Left: EIT~195~\AA\ image, middle: SXT image (AlMg filter), right: \ion{Mg}{xii} image. Images were taken on 1 October 2001.}
\label{F:mg_eit_sxt}
\end{figure}

The \ion{Mg}{xii} spectroheliograph was dedicated for hot plasma observations. It was launched in
2001 onboard the CORONAS-F satellite  \citep{ora02} as a part of the SPIRIT instrumentation complex
\citep{Zhitnik2002}. The instrument provides monochromatic solar coronal images in the \ion{Mg}{xii} 8.42~\AA\ resonance
line. The plasma emits this line at temperatures greater than 5~MK; therefore, the \ion{Mg}{xii} images contain only signal
from hot plasma without any low-temperature background. \ion{Mg}{xii} images differ from other telescopic images --- there is neither a solar limb nor a quiet Sun background (see Figure~\ref{F:mg_eit_sxt}). Typical structures in \ion{Mg}{xii} images range from 4 to 300~Mm in size and have a lifetime of several seconds up to several days \citep{Urnov07}. The \ion{Mg}{xii} spectroheliograph has two main advantages over `traditional' hot coronal imagers: 1) the entire signal shown in the \ion{Mg}{xii} images is from a hot plasma, and 2) faint features can be seen, because no cold background obscure them. 

In the present work, we used the \ion{Mg}{xii} spectroheliograph data obtained with a 105~s cadence and an $8^{\prime\prime}$
resolution. Due to instrumental effects, \ion{Mg}{xii} images are slightly elongated in one of the directions \citep{Reva2012}. We pre-processed the \ion{Mg}{xii} data by subtracting the  background and removing the signal caused by the radiation damage. To improve the visibility of faint features on the \ion{Mg}{xii} images, we used a power scale with index 0.3 (see Figures~\ref{F:mg_eit_sxt}~--~\ref{F:Mg_event1}). In the \ion{Mg}{xii} images of Figures~\ref{F:mg_eit_sxt}~--~\ref{F:Mg_event1}, blue and green correspond to low intensities, red and yellow to high intensities.

The MDI onboard the SOHO satellite \citep{Domingo1995}, maps the line-of-sight component of the photospheric magnetic field in a synoptic mode with a  $4^{\prime\prime}$ resolution and a 90~min cadence. 

The RHESSI observes HXR spectra from 3~keV up to 17~MeV. Using Fourier-based methods, RHESSI can synthesize HXR images in the same spectral range. RHESSI data were available for two out of four observed arcades. 

The EIT on SOHO satellite takes solar images at the wavelengths centered at 171, 195, 285, and 304~\AA. In a synoptic mode, EIT takes images in all four
channels every 6 h; in the `CME watch' mode the telescope takes images in the 195~\AA\ channel every 12~min. The
pixel size of the telescope is $2.6^{\prime\prime}$ and the spatial resolution is $5^{\prime\prime}$. We used `CME
watch' EIT data for auxiliary purposes~---~to observe  the cool plasma. Unfortunately, no $H_\alpha$ images were available for the analyzed arcades.

\section{Results}

\begin{figure}[t]
\centering
\includegraphics[width = \textwidth]{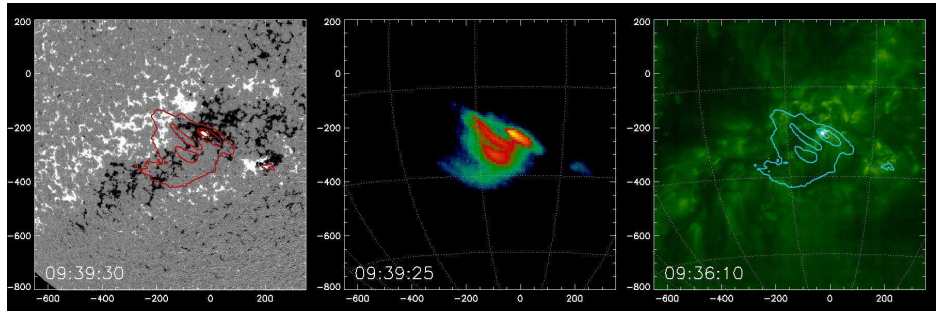}
\caption{The hot loop arcade of 28~February 2002 at 09:23 UT. Left: MDI image (the movie is available as electronic supplementary material); the red contours denote the \ion{Mg}{xii} signal. Middle: \ion{Mg}{xii} 8.42~\AA\ image. Right: EIT 195 \AA\  image; the blue contours denote the \ion{Mg}{xii} signal. The coordinates are in units of arcsec.}
\label{F:mdi_mg_eit}
\end{figure}

\begin{figure}[!t]
\centering
\includegraphics[width = \textwidth]{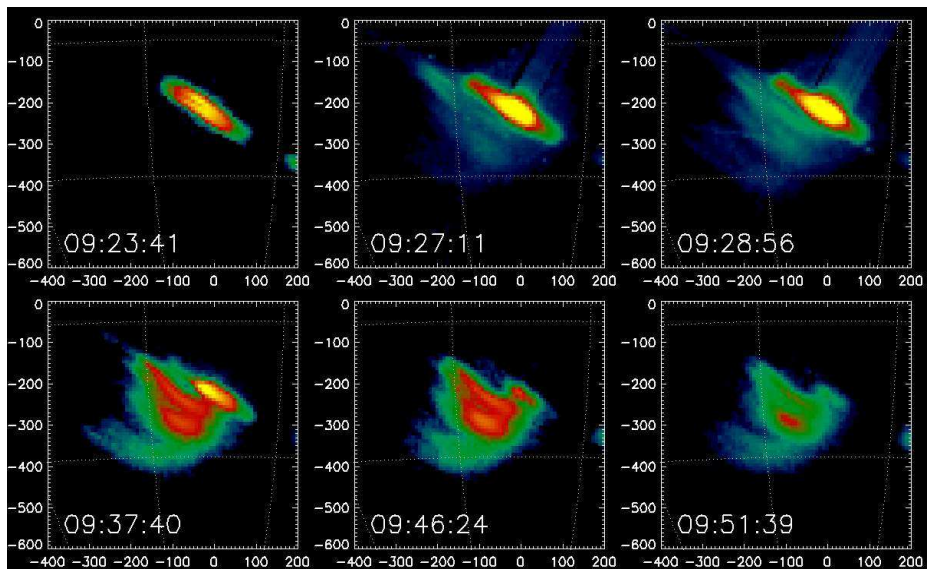}
\caption{The hot arcade of 28~February~2002 at 09:23 UT; `0' marks the precursor structure and `1'~--~`5' denote the arcade. The coordinates are in units of arcsec. (A movie is available as electronic supplementary material.)}
\label{F:Mg_event1}
\end{figure}

\begin{figure}[t]
\centering
\includegraphics[width = 0.49\textwidth]{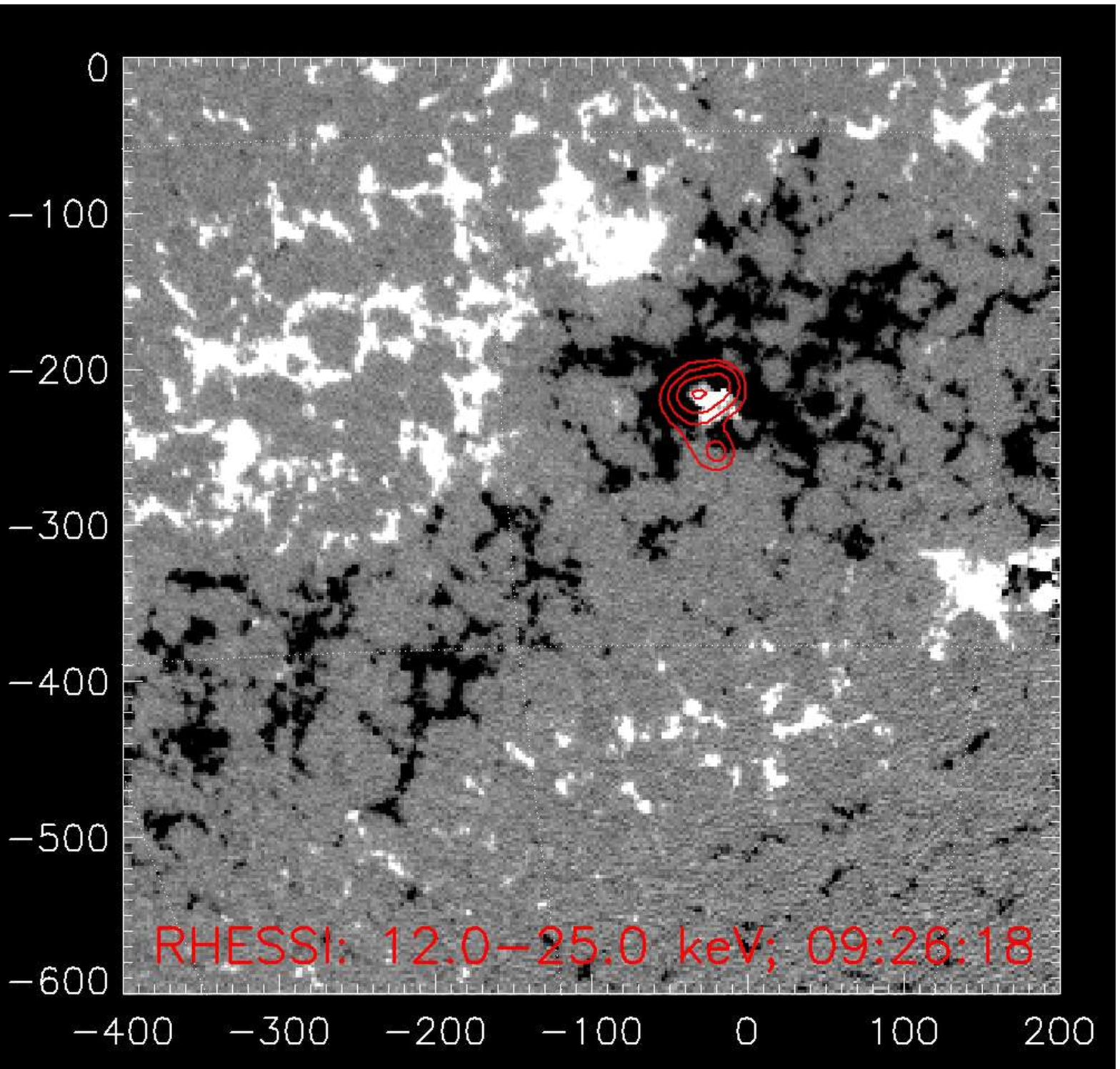}
\includegraphics[width = 0.49\textwidth]{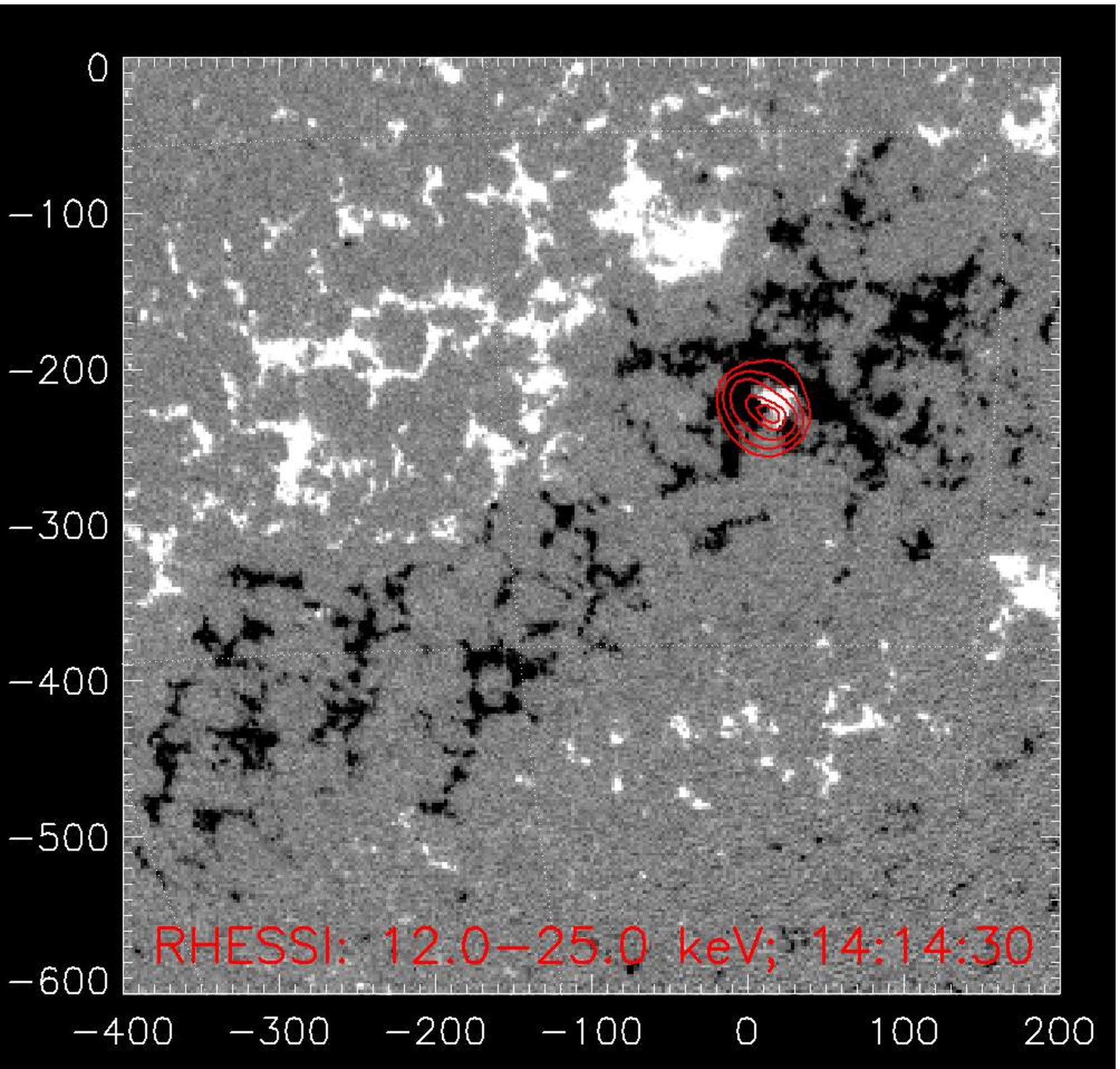}
\caption{RHESSI HXR sources (contours, reconstructed by the Pixon method) in the 12~--~25~keV spectral range superposed on the MDI magnetogram. The left panel shows the 09:26~UT event, the right panel the 14:14~UT event. There were no RHESSI data for the other two arcades. The coordinates are in units of arcsec.}
\label{F:RHESSI}
\end{figure}

All four hot arcades formed in the same active region (AR), above the polarity inversion line (PIL) between the two elongated regions of opposite polarities (AR NOAA 09847 and 09848; see Figure~\ref{F:mdi_mg_eit} left). On 25~February 2002, there was a flux emergence in the western part of the negative polarity region, and on 1~March 2002, it disappeared. This flux emergence caused 26~flares, and in four of them we observed the formation of hot loop arcades at 09:18, 14:13, and 22:28 UT on 28 February 2002, and at 00:40 UT on 1 March 2002 (see Figure~\ref{F:Mg_event1}). All four arcades evolved in the same way:

\begin{enumerate}
\item  A small hot source, which we call `the precursor', appeared above the flux emergence region in the \ion{Mg}{xii} images. At the same place and approximately at the same time, an HXR source appeared in the RHESSI images (see Figure~\ref{F:RHESSI}). This phase lasted for 2~--~10~min.
\item Then, a hot plasma was observed to fill the loops in a wave-like process: the loops that were closer to the precursor structure were filled earlier than the loops that were farther away. All four arcades formed in 5~min, which corresponds to a wave speed of approximately 700 km~s$^{-1}$.
\item The emission intensity of the loops increased and reached maximum after 5~–-~20~min. The maximum intensity was located in the apex of the loops above the PIL.
\item The emission intensity of the loops gradually decreased over approximately 1~h.
\end{enumerate}

Table~\ref{T:arcades} lists the characteristics of the observed arcades.

\begin{table}[!b]

\caption{Arcades parameters.}
\begin{tabular}{p{0.03\linewidth}p{0.13\linewidth}p{0.07\linewidth}p{0.11\linewidth}p{0.09\linewidth}p{0.1\linewidth}p{0.1\linewidth}p{0.09\linewidth}}
\hline 
No & Date        & GOES class &  Precursor start & First loop start & Last loop start & Arcade faded away & Number of loops\\ 
\hline
1  & 28 Feb 2002 & C4.0       &  09:20 & 09:27 & 09:29 & 11:20 & 5 \\ 
2  & 28 Feb 2002 & C5.5       &  14:13 & 14:16 & 14:20 & 15:45 & 3 \\ 
3  & 28 Feb 2002 & C7.5       &  22:35 & 22:40 & 22:45 & 23:49 & 3 \\ 
4  & 01 Mar 2002 & C3.0       &  00:37 & 00:42 & 00:49 & 01:33 & 4 \\ 
\hline 
\end{tabular} 
\label{T:arcades}
\end{table}

All four arcades had similar sizes; the loop length was 170~Mm, and the arcade length was 200~Mm (see Figure~\ref{F:mdi_mg_eit} center). The arcades consisted of 3~--~5 loops separated at $\approx 50$~Mm. The brightness of the loops gradually decreased with distance from the precursor structure. The loop footpoints were rooted in the main regions of positive and negative polarities (see Figure~\ref{F:Mg_event1}). 

In the EIT 195~\AA\ images, the precursor structure looked like a loop with a size of 20~Mm, but the arcade was invisible (see Figure~\ref{F:mdi_mg_eit} right). When the arcade cooled down and disappeared from the \ion{Mg}{xii} images, the brightest loop of the arcade appeared in the EIT 195~\AA\ images. 

\begin{figure}[hbtp]
\centering
\includegraphics[width = \textwidth]{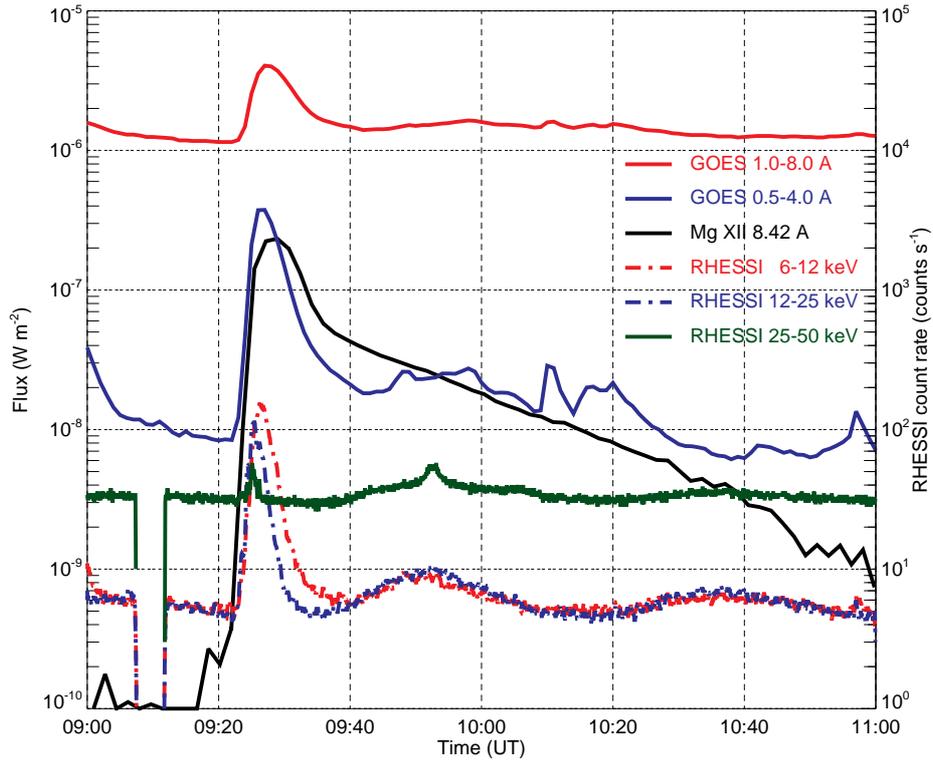}
\caption{Light curves of the arcade that occurred on 28 February 2002 at 09:23 UT; we indicate the 1~--~8~\AA\ (red) and  0.5~--~4~\AA\ (blue) channels of GOES, \ion{Mg}{XII}~8.42~\AA\ (black),  6~--~12~keV (red dash-dotted), 12~--~25~keV (blue dash-dotted), 25~--~50~keV (green) channels of RHESSI.}
\label{F:GOES}
\end{figure}

The HXR source appeared above the flux emergence region at 09:26 and 14:14~UT (events 1 and 2; see Figure~\ref{F:RHESSI}). There were no other HXR emissions during the events. RHESSI data were unavailable for the other two events. Figure~\ref{F:GOES} shows RHESSI, GOES, and \ion{Mg}{XII} light curves for the arcade that occurred at 09:26~UT. The majority of the flux in Figure~\ref{F:GOES} came from the precursor structure. The hard channels peaked earlier than the soft channels (RHESSI: 25~--~50~keV peaked at 09:25:04~UT, 12~--~25~keV at 09:25:32~UT, 6~--~12~keV at 09:26:16~UT; the GOES channels peaked at 09:27:00~UT; and \ion{Mg}{XII} peaked at 09:28:56~UT), which is consistent with the standard flare model. 

\begin{figure}[hbtp]
\centering
\includegraphics[width = 0.8\textwidth]{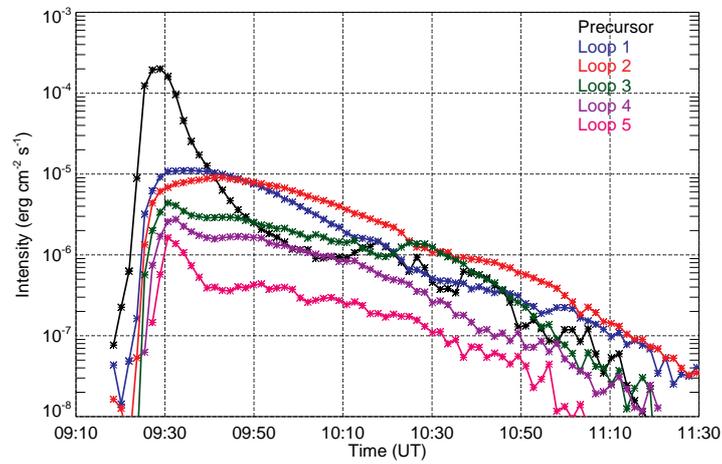}
\caption{Light curves of the precursor structure and individual loops of the arcade in the \ion{Mg}{xii} 8.42~\AA\ line. The arcade occurred after the flare on 28~February 2002 at 09:23 UT. The light curves of other arcades were similar to these curves.}
\label{F:LC1}
\end{figure}

\begin{figure}[hbtp]
\centering
\includegraphics[width = 0.7\textwidth]{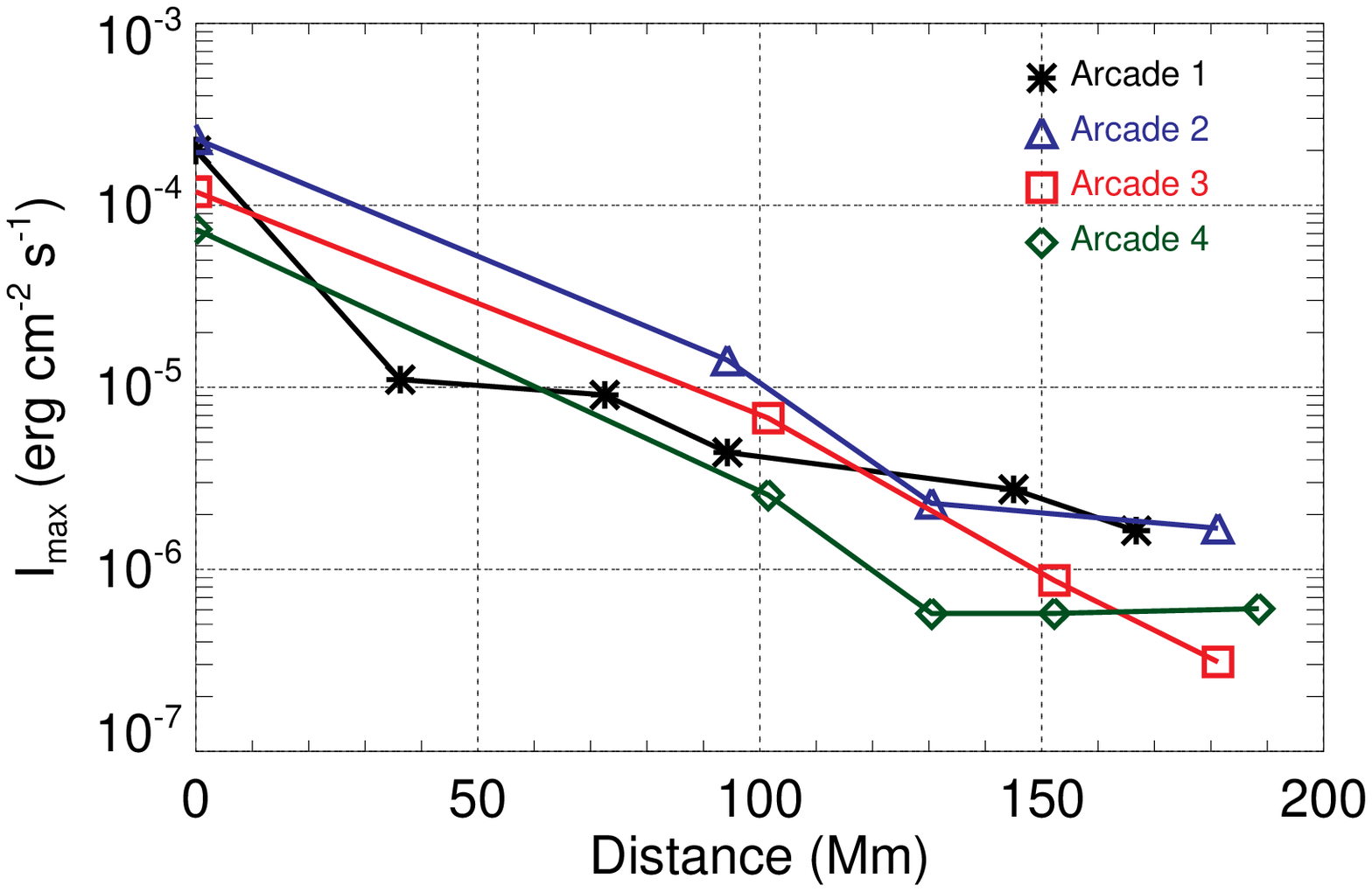}
\caption{Maximum intensity of the loops observed in the \ion{Mg}{xii} 8.42~\AA\ line as a function of the distance from the precursor structure.}
\label{F:Imax_Tmax}
\end{figure}

We measured the  light curves in the \ion{Mg}{xii} 8.42~\AA\ line of the precursor structures and individual loops of the arcades (see Figure~\ref{F:LC1}). The maximum intensity of the precursor structure exceeded the maximum intensity of the brightest loop by one order of magnitude and by two orders of the intensity of the faintest loop. The maximum intensity of the loops exponentially decreased with the distance from the precursor structure (see Figure~\ref{F:Imax_Tmax}). The intensity e-folding distance is $35 \pm 5$~Mm.

To measure the plasma cooling time ($\tau$), we approximated the light curves of the precursor structure and the loops by the following formula:
\begin{equation}
I = I_0 {\rm e}^{-t/\tau}.
\end{equation}
The cooling time was 4~--~6~min for the precursor structure and 15~--~35 min for the loops (see Table~\ref{T:Decay}). Although it is intuitive to interpret the longer lifetime of the loops as a sign of continuous heating, it is also possible to explain it as a pure cooling (see the Appendix).

\begin{table}[!b]
\caption{Decay times (measured in minutes) of the precursor structure and the loops of the arcade.}
\begin{tabular}{ccccccc}
\hline 
Event & Precursor & Loop 1 & Loop 2 & Loop 3 & Loop 4 & Loop 5 \\ 
\hline 
1     & 4.5       & 19     & 21     & 36     & 32     & 31     \\ 
2     & 6.6       & 21     & 24     & 24     & ---    & ---    \\ 
3     & 4.2       & 24     & 23     & 21     & ---    & ---    \\ 
4     & 5.4       & 16     & 19     & 24     & 10     & ---    \\ 
\hline 
\end{tabular} 
\label{T:Decay}
\end{table}

\section{Discussion}

\begin{figure}[t]
\centering
\includegraphics[width = 0.7\textwidth]{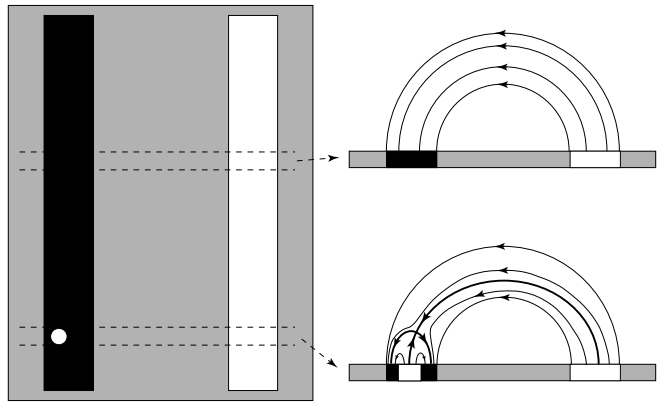}
\caption{Schematic magnetic field model of the event. Black and white indicate negative and positive polarities, and gray means neutral polarity. The quasi-vertical current sheet above the arcade is not shown here (see Figure~\ref{F:wave_model}).}
\label{F:Magnetic_model}
\end{figure}

Some aspects of the observed arcades are unusual; the first one is the magnetic configuration in which the events occurred. The precursor occurred above the flux emergence region in a quadrupolar magnetic configuration, suitable for flaring reconnection (see Figure~\ref{F:Magnetic_model}). It is clear that the flux emergence caused the precursor reconnection. However, the arcade loops formed in a simple bipolar configuration (see Figure~\ref{F:Magnetic_model}), without changes in their magnetic field structure.

Second, the arcades formed in a wave-like manner, which indicates that waves might play a role in the phenomena. Third, there was a cold space (distance) between the hot loops. It is unclear why some loops of the arcade were heated and some were not. Neither the wave-like formation nor the cold space can be explained by the standard 2.5D flare model. We need an extension of the standard flare model into 3D, which will explain the observations.

\begin{figure}[hbtp]
\centering
\includegraphics[width = \textwidth]{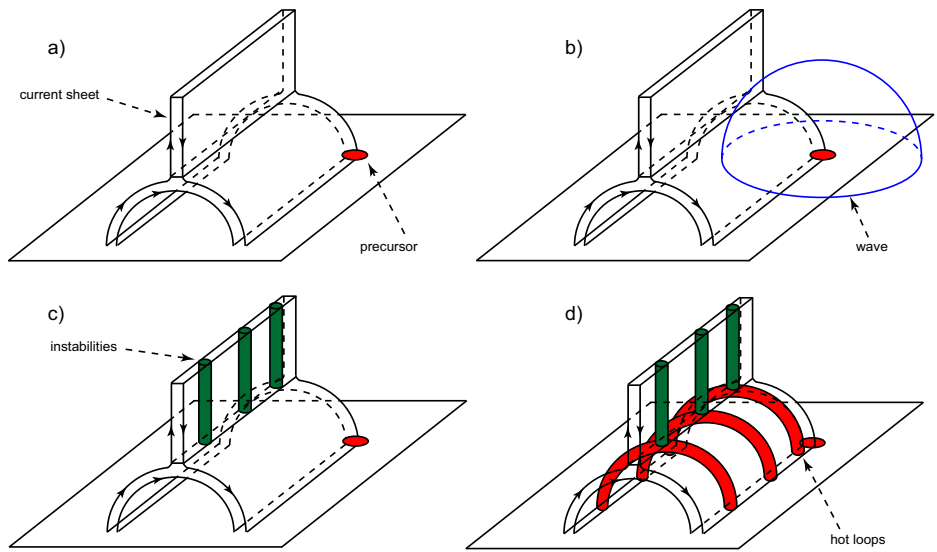}
\caption{Schematic model of hot loop arcade formation.}
\label{F:wave_model}
\end{figure}

To explain the observations, we propose that the current sheet existed above the loop apexes before the arcades
ignited (see Figure~\ref{F:wave_model}). We think that the arcade evolution was the following:
\begin{enumerate}
\item The precursor structure was formed at the edge of the arcade (see Figure~\ref{F:wave_model}a). 
\item The precursor launched an MHD wave, which propagated along the arcade (see Figure~\ref{F:wave_model}b). 
\item The MHD wave caused instabilities in the current sheet (see Figure~\ref{F:wave_model}c).  
\item The instabilities led to the heating of the underlying loops (see Figure~\ref{F:wave_model}d).
\end{enumerate}

\citet{Artemyev2012} studied the unstable wave modes of a quasi-vertical 2.5D current sheet above an arcade with a sheared magnetic field. The studied waves propagated in the current sheet along the PIL. The development of the symmetrical sausage-mode waves modes to the local thinning and thickenings of the current sheet.  Due to the conservation of magnetic flux, the current density is higher where the current sheet is thinner. This process intensifies the tearing instability \citep{Lapenta2000, Wiegelmann2000, Lapenta2003} and modulates the efficiency of the energy release along the current sheet. Thus, the propagation of the unstable wave modes can lead to successive episodes of plasma heating in different loops of the arcade. 

The characteristic distance between the two most effectively heated neighboring loops should approximately be equal to the wavelength ($\lambda$) of the corresponding unstable wave mode. The characteristic speed of the triggering front propagating along the PIL should correspond to the group speed of the wave ($v_{\rm g}$). The most appropriate unstable wave mode for the flaring loop arcades has the group speed \citep{Artemyev2012}

\begin{equation}
	v_{\rm g} \approx \frac{v_{\rm A}}{4} \sqrt{\frac{\lambda}{L_x} \sin{\varphi}},
	\label{eq1}
\end{equation}

where $v_{\rm A}$ is the Alfv\'en speed outside of the current sheet, $L_x$ is the
vertical scale of the current sheet, and $\varphi$ is the shear angle. 

Let us apply this theory to the observations. The impulsive energy release during the precursor process could initiate the unstable wave modes in the current sheet above the arcades. \citet{Artemyev2012} predicted that the wavelength lies in the range from 0.1~Mm to 1000~Mm. The observed value $\lambda \approx 50$~Mm corresponds to the collisionless mode 3 found in \citet{Artemyev2012}. From the observations, we estimate that $v_{\rm g} \approx 700$~km~s$^{-1}$ and $\varphi \approx 30^\circ$ (see
Figure~\ref{F:mdi_mg_eit}). Thus, to satisfy Equation~(\ref{eq1}), $L_x$ should be

\begin{equation}
	L_x \approx 3 \times 10^{-3} \, v^2_{\rm A} \quad \mbox{km}
\end{equation}

Here, the Alfv\'en speed $v_{\rm A}$ is in km~s$^{-1}$. For the estimation, we used the `canonical' value for the solar corona
($v_{\rm A}=1000$~km~s$^{-1}$) and found that in the studied events $L_x \approx 3$~Mm.  This length is similar to what was found by \citet{Reid2011} in their analysis of the coronal type III radio bursts. Finally, the estimated vertical length of the current sheet is 6~--~7 orders of magnitude greater than the
characteristic ion gyroradius in the solar corona (current sheet width), which is  a favorable
condition for magnetic reconnection and plasma heating. 

Let us discuss other possible models. \citet{Emslie1981} proposed that an increase in plasma pressure can increase the cross section of a hot flare loop. The hot loop will interact with a neighboring cold loop of the arcade and cause magnetic reconnection and heating in the cold loop. This process can propagate along the arcade like a domino effect. However, this mechanism cannot explain the cold space between the hot loops.

\citet{Nakariakov2011} proposed that a slow magnetoacoustic wave that propagate along the arcade can trigger energy release in the current sheet above the arcade. The main difference with the model of \citet{Artemyev2012} is the nature of the wave. In the model of \citet{Artemyev2012} the wave propagates inside the current sheet, while in the model of \citet{Nakariakov2011}  the wave propagates along the arcade outside the current sheet. In our observations, a group of slow magnetoacoustic waves could be initiated by the precursor. The model of \citet{Nakariakov2011} can explain the wave-like formation of a hot arcade, but it cannot easily explain the cold space between the hot loops.

\citet{Liu2009} analyzed the flares in which a filament erupted asymmetrically along the PIL. In these events, the arcade should brighten in a wave-like manner. Although filaments can erupt several times in a row from the same active region \citep{Archontis2014}, the model fails to explain the cold space between the hot loops. 

The model of \citet{Artemyev2012} explains both the wave-like formation of the arcades and the cold space between the hot loops. The model gives a reasonable estimate of the current sheet vertical scale. The observed distance between the hot loops is consistent with the model. The \citet{Artemyev2012} model fits our observations better than other models known to us. We hope that the reported observations will stimulate new theoretical investigations and observations of this type of solar flare phenomenon.

\section{Conclusions}

We presented the first observations of the formation of hot loop arcades in the hot monochromatic line. Thanks to the relatively high cadence of the \ion{Mg}{xii} spectroheliograph, we saw the formation of the arcades in detail. At first, a small source of soft X-ray and hard X-ray emissions appeared at the edge of the future arcade. Then the arcade brightened in a wave-like manner. The wave speed was about 700~km~s$^{-1}$, which is the same order of magnitude as the MHD coronal wave speed. The arcades occurred four times in a row in the same place under the same conditions, following the same scenario.

We interpreted the observations in terms of the MHD waves. We think that the current sheet existed above the loop apexes before the arcade ignited. The precursor launched an MHD wave, which triggered instabilities in the current sheet. The instabilities led to the reconnection and loop heating.

Our interpretation is consistent with previous theoretical works \citep{Somov1982, Artemyev2012}. The observations show that flares are essentially a 3D processes 2D models do not explain all their aspects. The observations also point out that MHD waves could play an important role in flare processes; they could transfer energy and trigger reconnection.

\acknowledgments
We are grateful to Boris Somov and Anton Artemyev for their invaluable help. This work was  supported by a grant from the Russian Foundation of
Basic Research (grant 14-02-00945) and by the Program No. 22 for fundamental
research of the Presidium of the Russian Academy of Sciences.

\appendix
\section*{Cooling times}

The loops of the arcade kept their high temperature for a long time (see Figure~\ref{F:LC1}). This could be interpreted as a sign of the continuous loop heating. To verify this conclusion, we here estimate the plasma cooling time in the absence of heating, and compare it with the measured values. 

There are two mechanisms of loop cooling: conductive and radiative. At high temperatures ($\approx$~10~MK), the conductive cooling dominates radiative cooling. To estimate the
conductive cooling time ($\tau_{\rm cond}$), we use the formula \citep{cul94}
\begin{equation}
\tau_{\rm cond} = \frac{21 n_{\rm e} k_{\rm B} L^2}{5 \kappa T^{5/2}},
\label{E:t_cond}
\end{equation}
where $\kappa = 9.2 \times 10^{-7} \ \textmd{erg} \ \textmd{s}^{-1}  \textmd{cm}^{-1}  \textmd{K}^{-7/2}$ is the Spitzer conductivity, $n_{\rm e}$ is the electron density, $k_{\rm B}$ is the Boltzmann constant, $L$ is the loop length, and $T$ is the loop temperature.

\begin{figure}[t]
\centering
\includegraphics[width = 0.6\textwidth]{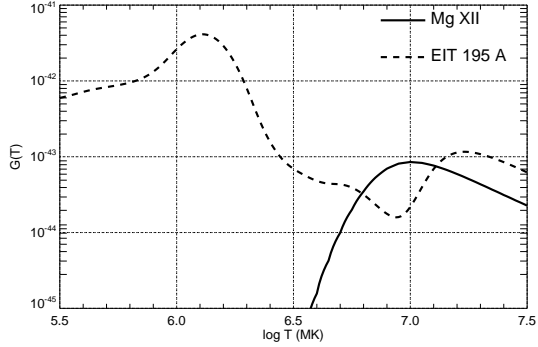}
\caption{Temperature response functions of the \ion{Mg}{xii} spectroheliograph (solid line) and EIT 195~\AA\ telescope (dashed line).}
\label{F:GofT}
\end{figure}

To estimate the arcade temperature, we compared \ion{Mg}{xii} and EIT 195~\AA\ images. The EIT 195~\AA\ channel
is sensitive to 1~MK and 16~MK plasma (see Figure~\ref{F:GofT}). The \ion{Mg}{xii} spectroheliograph is sensitive to
plasmas hotter than 5~MK (see Figure~\ref{F:GofT}). Since we see the precursor structure in both EIT 195~\AA\ and
\ion{Mg}{xii} channels, the precursor temperature should be about 15~MK. Since we were able to see the  arcade in the
\ion{Mg}{xii} line but not in the EIT images, the loop temperature should lie in the range of 5~--~10~MK. 

To estimate the precursor electron density, we estimated its temperature ($T$) and emission measure (EM) under isothermal approximation using the filter ratio method with EIT~195~\AA\ and \ion{Mg}{xii} fluxes. The result is $T \approx$~14~MK and EM $\approx 8.8 \cdot 10^{46}$~cm$^{-3}$. The precursor structure had a length of $L \approx 20$~Mm and the width of the EIT pixel. However, since loops in the AIA images have the width of the AIA pixel ($r$~=~0.44~Mm), we think that the AIA pixel size is a more reasonable estimate for the loop width than the EIT pixel size. Now, we can estimate the precursor electron density:

\begin{equation}
{\rm EM} = n_{\rm e}^2 \pi r^2 L \Rightarrow n_{\rm e} = \sqrt{\frac{\rm EM}{\pi r^2 L}} \approx 8.6 \cdot 10^{10} \ \mbox{cm}^{-3}.
\end{equation}

Unfortunately,  we have only \ion{Mg}{xii} observations for the flaring loops, so that there is no way to measure the emission measure and temperature of their plasma separately using the filter ratio. However, we can estimate their electron density from

\begin{equation}
I = G(T) n_{\rm e}^2 \pi r^2 L \leq G_{\rm max} n_{\rm e}^2 \pi r^2 L,
\end{equation} 
namely,
\begin{equation}
n_{\rm e} \geq \sqrt{\frac{I}{G_{\rm max} \pi r^2 L}} \approx 1.8 \cdot 10^{10} \ \mbox{cm}^{-3},
\end{equation}

where $I$ is the emission intensity of the loops on the \ion{Mg}{xii} images, $G(T)$ is the temperature response function of the \ion{Mg}{xii} spectroheliograph, and $G_{\rm max}$ is the maximum value of $G(T)$. For the sake of estimation, we used $n_{\rm e} = 10^{10}$~cm$^{-3}$ and $T$~=~10~MK for the loops of the arcade.

For the loops, we obtained $\tau_{\rm cond}$~=~100~min ($n_{\rm e} = 10^{10}$~cm$^{-3}$, $T$~=~10~MK, and $L$~=~170~Mm).
This is of the same order of magnitude as the measured values, 20~--~30 min. Furthermore, slight changes of the $n_{\rm e}$ and
$T$ values can make the agreement better. This means that it is possible that the loop heating was impulsive and long lifetime of the loops was due to their large size.

For the precursor, we obtained $\tau_{\rm cond}$~=~5~min ($n_{\rm e} = 8.6 \cdot 10^{10}$~cm$^{-3}$, $T = 14$~MK, and $L$~=~20~Mm), which coincides with the observed values. This means that it is possible that the heating of the precursor structure was impulsive.

It may be surprising that the long decay of the \ion{Mg}{xii} light curves can be interpreted as pure cooling. However, our estimate is very rough; it is valid in the temperature range $6.8 < \log T < 7.3$, where the response function of \ion{Mg}{xii} channel  does not change much. Therefore, pure cooling is only an option, but not a fact. 

Furthermore, other mechanisms could decrease the loop intensity; for example by mass draining. The mass draining could play a significant role in the late phase of flare decay \citep{Bradshaw2010b}. The draining will decrease the electron density, and therefore the loop intensity will decrease faster. Also, according to Equation~(\ref{E:t_cond}), the loops will cool faster. If the draining was present, then the hot loops would require continuous heating to support their long decay times. 

\bibliographystyle{spr-mp-sola}
\bibliography{mybibl}

\end{article}
\end{document}